\documentclass[sigconf, nonacm, pdfa]{acmart}

\usepackage{balance}
\usepackage{cite}
\usepackage{amsmath,amsfonts}
\usepackage{algorithmic}
\usepackage{booktabs}
\usepackage{graphicx}
\usepackage{textcomp}
\usepackage{hyperref}
\usepackage{enumitem}
\usepackage{multirow}
\usepackage{subcaption}
\usepackage{framed}
\usepackage{float}
\usepackage{booktabs}
\usepackage{tikz}
\usepackage{xcolor,colortbl}
\usepackage{pifont}
\newcommand{\xmark}{\ding{55}}

\newcommand\vldbdoi{10.14778/3742728.3742761}
\newcommand\vldbpages{2735 - 2747}
\newcommand\vldbvolume{18}
\newcommand\vldbissue{8}
\newcommand\vldbyear{2025}
\newcommand\vldbauthors{\authors}
\newcommand\vldbtitle{\shorttitle} 
\newcommand\vldbavailabilityurl{https://github.com/yeounoh/lc_nl2sql}
\newcommand\vldbpagestyle{empty}

\definecolor{mycolor}{rgb}{0.97, 0.91, 0.81}

\newcommand{\yeounoh}[1]{
#1
}

\usepackage[a-2b]{pdfx}
\begin{document}


\title{Is Long Context All You Need? Leveraging LLM's Extended Context for NL2SQL}

\settopmatter{authorsperrow=5}

\author{Yeounoh Chung}
\affiliation{
 \institution{Google}
 \country{}
}
\email{yeounoh@google.com}
\author{Gaurav T. Kakkar}
\affiliation{
 \institution{Google}
 \country{}
}
\email{gkakkar@google.com}
\author{Yu Gan}
\affiliation{
 \institution{Google}
 \country{}
}
\email{gany@google.com}
\author{Brenton Milne}
\affiliation{
 \institution{Google}
 \country{}
}
\email{bmil@google.com}
\author{Fatma \"{O}zcan}
\affiliation{
 \institution{Google}
 \country{}
}
\email{fozcan@google.com}

\begin{abstract}
Large Language Models (LLMs) have demonstrated impressive capabilities across a range of natural language processing tasks. In particular, improvements in reasoning abilities and the expansion of context windows have opened new avenues for leveraging these powerful models. 
NL2SQL is challenging in that the natural language question is inherently ambiguous, while the SQL generation requires a precise understanding of complex data schema and semantics. One approach to this semantic ambiguous problem is to provide more and sufficient contextual information. 

In this work, we explore the performance and the latency trade-offs of the extended context window (a.k.a., long context) offered by Google's state-of-the-art LLM (\textit{gemini-1.5-pro}). 
We study the impact of various contextual information, including column example values, question and SQL query pairs, user-provided hints, SQL documentation, and schema. To the best of our knowledge, this is the first work to study how the extended context window and extra contextual information can help NL2SQL generation with respect to both accuracy and latency cost.
We show that long context LLMs are robust and do not get lost in the extended contextual information. Additionally, our long-context NL2SQL pipeline based on  Google's \textit{Gemini-pro-1.5} \yeounoh{achieves 
strong performance across multiple benchmark datasets} 
without fine-tuning or expensive self-consistency based techniques. 
\end{abstract}
\settopmatter{printacmref=false, printfolios=true}

\maketitle

\pagestyle{\vldbpagestyle}
\begingroup\small\noindent\raggedright\textbf{PVLDB Reference Format:}\\
\vldbauthors. \vldbtitle. PVLDB, \vldbvolume(\vldbissue): \vldbpages, \vldbyear.\\
\href{https://doi.org/\vldbdoi}{doi:\vldbdoi}
\endgroup
\begingroup
\renewcommand\thefootnote{}\footnote{\noindent
This work is licensed under the Creative Commons BY-NC-ND 4.0 International License. Visit \url{https://creativecommons.org/licenses/by-nc-nd/4.0/} to view a copy of this license. For any use beyond those covered by this license, obtain permission by emailing \href{mailto:info@vldb.org}{info@vldb.org}. Copyright is held by the owner/author(s). Publication rights licensed to the VLDB Endowment. \\
\raggedright Proceedings of the VLDB Endowment, Vol. \vldbvolume, No. \vldbissue\ %
ISSN 2150-8097. \\
\href{https://doi.org/\vldbdoi}{doi:\vldbdoi} \\
}\addtocounter{footnote}{-1}\endgroup

\ifdefempty{\vldbavailabilityurl}{}{
\vspace{.3cm}
 \begingroup\small\noindent\raggedright\textbf{PVLDB Artifact Availability:}\\
 The source code, data, and/or other artifacts have been made available at \url{\vldbavailabilityurl}.
\endgroup
}

\section{Introduction}

Recent advancements in LLMs have particularly focused on enhancing their retrieval and reasoning abilities and expanding their context windows, thus broadening the scope of their potential applications. The ability to process and retain longer sequences of information within the context window empowers LLMs to capture nuanced dependencies and intricate relationships within the input data, offering unprecedented possibilities for improved language understanding and generation. One domain that stands to significantly benefit from these advancements is Natural Language to SQL (NL2SQL). NL2SQL is a challenging task that entails translating natural language questions into structured SQL queries that can be executed on a database. The inherent ambiguity of natural language questions, coupled with the necessity for a deep understanding of complex database schemas and semantics, makes NL2SQL a very challenging problem~\citep{Floratou2024NL2SQLIA}. 
Recent works \citep{pourreza2024chase,maamari2024death,caferouglu2024sql,xiyansql} created NL2SQL pipelines involving schema linking, self-consistency, and self-correction with multiple LLM calls. These solutions carefully create prompts that include various context information, and use Chain-of-Thought (CoT)\citep{liu2023divide,zhou2022least} reasoning and/or in-context-learning. 

In this paper, we explore the potential of harnessing the extended context window provided by Google's long context LLMs (\textit{gemini-1.5}) to improve NL2SQL performance.
\yeounoh{The hypothesis is that the long-context LLMs with enhanced retrieval and reasoning abilities over this extended context can address the semantic ambiguity challenges with additional and appropriate contextual information.
To study this hypothesis}, we conduct a detailed study on the impact of these techniques when used with a long-context model.  


\begin{table}[t!]
\centering
\begin{minipage}{\columnwidth}
\caption{\small Performance comparison of different published NL2SQL methods on BIRD dev. This excludes undisclosed methods from the leaderboard.}\vspace{-3mm}
\label{table:bird_dev_results}
\resizebox{\textwidth}{!}{
    \begin{tabular}{lccc}
    \toprule
    Method & Ex Acc (\%) & Fine-tuned & Self-consistency \\
    \midrule
    CHASE-SQL\footnote[1]{2\textsuperscript{nd} on the leaderboard as of December 17th, 2024. Long context NL2SQL pipeline was used as one of the NL2SQL pipelines to generate multiple output candidates} \citep{pourreza2024chase} &  74.46 & \checkmark & \checkmark \\
    \yeounoh{XiYan-SQL} &  73.34 & \checkmark & \checkmark \\ 
    \textbf{Long Context (ours)} &  \textbf{67.41} & \xmark & \xmark \\
    Distillery\footnote[3]{6\textsuperscript{th} on the leaderboard as of December 17th, 2024}\citep{maamari2024death}  & 67.21 & \checkmark & \checkmark \\
    E-SQL\citep{caferouglu2024sql} & 65.58 & \xmark & \xmark \\
    CHESS\citep{talaei2024chess} & 65.00  & \checkmark & \checkmark \\
    MCS-SQL\citep{lee2024mcs} & 63.36 & \xmark & \checkmark \\
    SuperSQL\citep{li2024dawn} & 58.5 & \xmark & \checkmark \\
    \bottomrule
    \end{tabular}
}
\end{minipage}%
\vspace{-0.5cm}
\end{table}

Table~\ref{table:bird_dev_results} compares the execution results accuracy (Ex Acc) of published NL2SQL methods \yeounoh{using BIRD benchmark~\citep{li2024can}, the most popular benchmark for NL2SQL testing.} Our \textit{Long Context} pipeline yields very competitive results without any fine-tuning and without generating multiple answer candidates (self-consistency). 
\yeounoh{It is shown that LLMs, especially smaller capacity models, benefit significantly from fine-tuning as it helps the models focus on relevant patterns in specific data domains and SQL generation~\citep{dominguez2024blar,liu2024survey,zhou2024dbgpthub}. 
While fine-tuning has been a dominant approach, in-context learning (ICL) using the latest LLMs is gaining traction as a preferable alternative, matching the performance of fine-tuned models~\citep{nan2023enhancing,pourreza2024din}. ICL does not require re-training or updating the model parameters and avoids overfitting to specific data domains. This is particularly appealing for a production NL2SQL system where it serves many customers across many specialized domains.
}

\yeounoh{
Self-consistency is another popular technique used in state-of-the-art NL2SQL systems~\citep{wang2022self,talaei2024chess,pourreza2024chase,dong2023c3,dail_sql,xiyansql}. The idea is to generate multiple output candidates and pick the most likely one according to some rules (e.g., majority voting) or a fine-tuned picker model. This works nicely due to the stochastic nature of LLMs, where different models (and even the same model) generate varying outputs for the same input.
Self-consistency is orthogonal to our work, and one can combine the two~\citep{pourreza2024chase} for higher accuracy (see Section~\ref{sec:self_consistency} for more discussion). In this work we focus on identifying and providing additional contextual information for ICL, leveraging the extended context window of a long-context LLM.
 }

\yeounoh{
\textit{E-SQL} is another method that does not employ fine-tuning or self-consistency. Instead,} \textit{E-SQL} focuses on improved mapping between user question and the relevant schema elements (a.k.a., schema linking).
\yeounoh{It modifies/enriches the original question explicitly with relevant schema elements (table and column names, values) and conditions, so the model can avoid the implicit mapping of question to relevant schema elements during generation. Such techniques to improve schema linking, just like fine-tuning and self-consistency, are orthogonal and complementary to our work.} 

In this paper,  we conduct extensive empirical evaluations on established NL2SQL benchmarks, such as the BIRD and SPIDER datasets, to investigate the impact of long context on NL2SQL performance. We explore various techniques for effectively leveraging long context in the NL2SQL pipeline, including strategies for context construction, prompt engineering, and agentic workflow creation. \yeounoh{To the best of our knowledge, this is the first study exploring the real potential and implications of utilizing a long context LLM and its extended context window of millions of tokens for NL2SQL, yielding competitive performance on popular benchmarks. Google \textit{gemini-1.5} is currently the only long-context LLM that supports millions of tokens, whereas other models support up to 8k - 128k tokens in the context. Due to this constraint, prior works focused on squeezing the filtered relevant information onto limited context sizes.} Our findings demonstrate that long context can indeed serve as a powerful tool for enhancing a typical LLM-based NL2SQL pipeline without fine-tuning. By employing Google's \textit{gemini-pro-1.5} and its long context capabilities, our NL2SQL pipeline achieves 67.41\% accuracy on the BIRD-Bench dev dataset, showcasing the potential of this approach. Through detailed analyses, we provide insights into how long context enhances NL2SQL generation, highlighting the specific types of contextual information that are most beneficial. In particular, we observe the following insights:
\begin{itemize}
    \item  We observe that having the correct table and columns, i.e., 100\% recall, in the context is required for high quality SQL generation. The long context model does not get distracted by additional table information, i.e., when we have a large number of irrelevant tables in the context (low precision)
    \item  Our experiments show that simply adding many relevant examples (by question similarity) selected from \yeounoh{any available example pools (e.g., training dataset) does not necessarily} improve NL2SQL accuracy significantly, contrary to what we expected from the many-shot-learning paper~\citep{agarwal2024manyshot}. Instead, examples that are constructed using similar SQL constructs -- as the underlying dev dataset (but without looking at them) -- and relevant schema elements from the same target database improve accuracy. With this observation, we generate many such examples synthetically for many-shot ICL for NL2SQL. \yeounoh{It is also the first study of many-shot ICL for NL2SQL with hundreds of examples, whereas prior ICL approaches for NL2SQL utilize only a handful (3-5) of demonstrations/examples}.
    \item In our ablation study, hints have the highest impact on NL2SQL accuracy, followed by column sample values, and self-correction. While high-quality in-context examples improve accuracy, they do not solve all problems in NL2SQL generation. 
    \item \yeounoh{Our ablation studies show that different query types benefit from different contextual information.  Many-shot ICL with synthetic examples works for easy questions but not for the challenging BIRD dev questions.  On complex datasets like BEAVER dw (enterprise data warehouse), which involves more joins and complex SQL, synthetic examples actually hurt performance on average.}
    \item  Providing SQL documentation in the long context does not improve accuracy much, as the model has already seen these documents during training.
    \item Latency increases (near) linearly with context size, hence there is a clear trade-off between latency and better accuracy. 
\end{itemize}


In the remainder of this paper, we present a comprehensive overview of related work in Section \ref{sec:related}, followed by a detailed description of our long-context based NL2SQL approach in Section\ref{sec:approach}.  In  Section \ref{sec:experiments} we provide a detailed  analysis of various techniques and their impact on NL2SQL accuracy, as well as an ablation study.  Finally, we conclude with a discussion of the implications of our findings and directions for future research.

\section{Related Works}
\label{sec:related}
Early approaches in NL2SQL focused on rule-based and semantic parsing methods \citep{yin2020tabert,li2023resdsql}. 
With the advent of deep learning, neural network-based models have become dominant in NL2SQL research \citep{zhong2017seq2sql,wang-etal-2020-rat}.  These models learn to map natural language questions to SQL queries using large datasets of question-query pairs. These models treat NL2SQL as a machine translation task, using encoder-decoder architectures to translate natural language questions into SQL queries. More recently, LLMs with human-like abilities to understand and generate text have led to significant improvements in accuracy and efficiency in NL2SQL and similar tasks \citep{dail_sql,zan2022large,DBLP:journals/vldb/KatsogiannisMeimarakisK23,DBLP:journals/pvldb/Katsogiannis-Meimarakis23}. 

LLMs can also learn from a few examples provided in context at inference (few-shot in-context learning).  In the NL2SQL domain, prior research \citep{nan2023enhancing,pourreza2024din} focused on leveraging this few-shot in-context learning (ICL) approach to guide the SQL generation. The few-shot examples for NL2SQL consist of question and SQL pairs and are typically filtered by question embedding similarity to fit within the limited context size. In other problem domains, many-shot ICL enabled by the newly expanded context window is shown to consistently outperform the few-shot approach \citep{agarwal2024manyshot}. 

A new long context benchmark \citep{lee2024can} shows that the latest long-context LLMs can match the capability of state-of-the-art retrieval systems, while under-performing on compositional reasoning tasks, like NL2SQL against specialized fine-tuned models. This benchmark study covers NL2SQL and uses un-tuned \textit{gemini-1.5-pro} long context LLM. They leveraged the long context by including all the DB tables and also fixed few-shot examples as in \citep{dail_sql}. While their NL2SQL pipeline under-performs against the original \citep{dail_sql} with fine-tuned model on SPIDER 1.0 benchmark\citep{Yu.al.18c}, our long-context strategy outperforms both \citep{lee2024can,dail_sql}, by better leveraging the long context with appropriate information. 
\citep{bai2023longbench,li2024long} also studied ICL with long context for other non-NL2SQL domains. These studies show that many commercial long context LLMs struggle with processing all the information presented in the long context, where there is a bias towards certain locations (e.g., the end of window) \citep{liu-etal-2024-lost}. Our experiences point to the opposite direction where \textit{gemini-1.5-pro} exhibits no such strong bias and over much larger context window. As time of this writing, Google Cloud  \textit{gemini-1.5-pro} supports up to 2-million tokens in context over Vertex AI API, whereas OpenAI \textit{GPT-4o} supports 8k - 128k tokens over API.

Schema linking is critical for accurate NL2SQL generation \citep{Floratou2024NL2SQLIA}, as it maps ambiguous user questions to relevant schema elements. Schema linking often relies on careful selection of relevant tables and columns \citep{caferouglu2024sql,talaei2024chess}, and a recent study \citep{maamari2024death} showed that the latest LLMs can retrieve relevant schema elements from unfiltered database schema (i.e., passing all DB tables without selection) during inference. We also include the entire database schema in the long context and observe a similar trend. 

One can refine the output through multiple LLM calls, verifying, fixing and rewriting the SQL outputs. It is also common to sample multiple answer candidates and select (self-consistency) the most probable or consistent output to improve the quality \citep{wang2022self,talaei2024chess,pourreza2024chase,dong2023c3,dail_sql,xiyansql}.  In this work, we focus on leveraging the long context ICL for a typical NL2SQL pipeline -- which is orthogonal and can be used with self-consistency. \citep{pourreza2024chase} uses our long context pipeline to generate candidates along with a fine-tuned picker.

We do not use more complex prompting strategy, like Chain-of-Thought (CoT)\citep{wei2022chain,tai2023exploring,liu2023divide,zhou2022least}, as its contribution is negligible (see Section~\ref{sec:ablation} for more discussion). CoT prompting is shown to improve the performance of LLMs on arithmetic and commonsense reasoning tasks. It often involves providing a few demonstrations of intermediate reasoning steps to guide the model in generating its own reasoning chains.

\section{leveraging long context for NL2SQL}\label{sec:approach}

\begin{figure}[htbp]
\centering 
\includegraphics[width=\columnwidth]{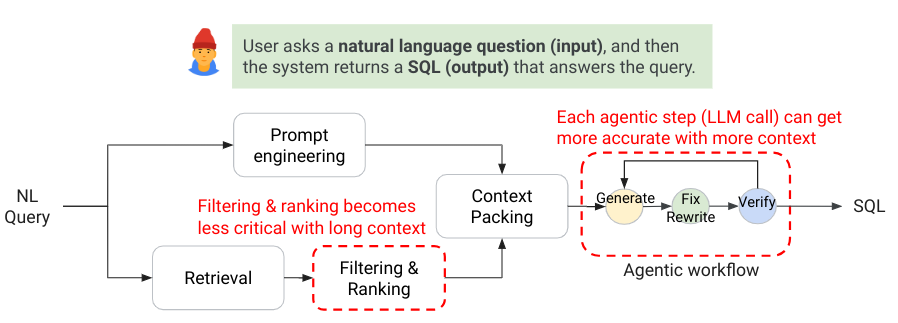} 
\caption{\textit{Long Context} NL2SQL Pipeline. Leveraging long context LLMs can make the retrieval step less critical and the agentic workflow (\textit{generate} $\rightarrow$ \textit{fix \& rewrite} $\rightarrow$ \textit{verify}) more accurate with extra contextual information.}
\label{fig:long_context_pipeline}
\vspace{-0.5cm}
\end{figure}
In this work, we focus on the additional contextual information that we can pass to the extended context window. We hypothesize that leveraging the extended context window and long-context LLMs' strong retrieval capability can make the retrieval step in NL2SQL pipeline less critical; we can also make LLM agent calls to generate, fix \& re-write, and to verify more accurate. We explore various contextual information, like SQLite documentation, user hints for clarification, extra column sample values and examples for many-shot in-context learning (ICL). 

This section is divided into three sub-sections, \textit{generate} $\rightarrow$ \textit{fix \& rewrite} $\rightarrow$ \textit{verify}, corresponding to the three agentic workflow calls for generating and refining the output SQL, shown in Fig.~\ref{fig:long_context_pipeline}. Each sub-section details the appropriate contextual information and techniques leveraging the extended context window per the corresponding LLM agent call. 

\subsection{Generate}
Focusing on the value of information rather than squeezing into the limited context window, we can provide a lot more contextual information to address challenges in schema linking and semantic errors stemming from ambiguous user question and complex data schema. We explore the following to assist the SQL generation.
\\\\
\noindent\textbf{\textit{All database tables for high recall schema linking.}}

In a typical NL2SQL pipeline, the relevant schema elements (table column names and values) are retrieved based on question embedding similarity. Accurate schema linking is critical for accurate SQL generation \citep{Floratou2024NL2SQLIA}. Prior research works focus on improving schema linking accuracy via more accurate table and column selection  \citep{caferouglu2024sql,talaei2024chess}. Accurate schema retrieval can uplift the NL2SQL performance, but there is still a lot of headroom (see Appendix~\ref{appendix:impact_col_selection} for more discussion). In fact, table schema retrieval can have missing tables and columns -- which prevents LLMs from generating correct SQL queries.

\begin{figure}[t!]
    \centering
        \centering
        \includegraphics[width=0.90\columnwidth, height=0.5\columnwidth]{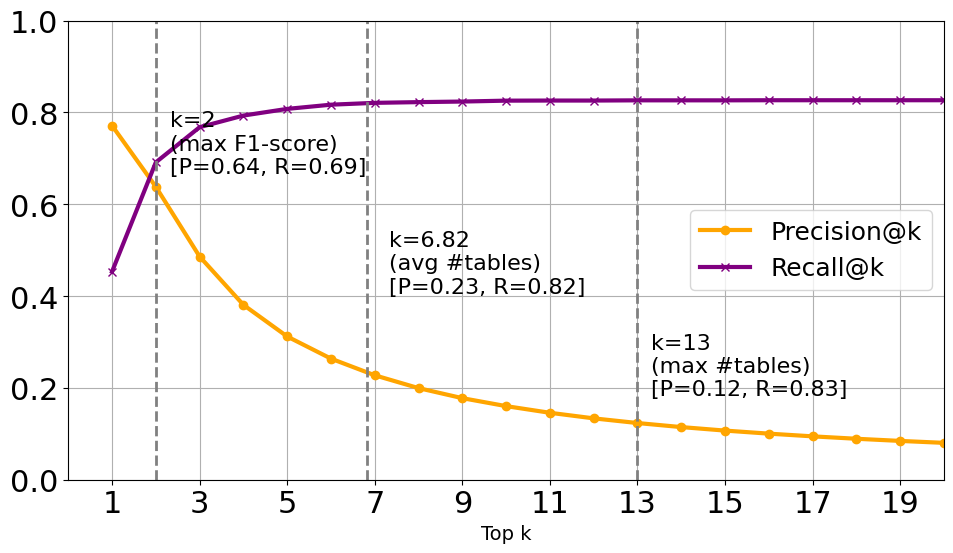} 
    \caption{Top K relevant table retrieval simulation. The top-K relevant tables are selected from BIRD dev based on question embedding similarity, \yeounoh{without a specific target DB, as supported in some production RDBMSs.}}
    \label{fig:tbr_simulation}

\end{figure}

Figure~\ref{fig:tbr_simulation} illustrates table schema retrieval simulation results mimicking a production table retrieval setup. Note that the recall does not reach 100\% and plateaus at around 82\% with increasing $K$.
\yeounoh{In some production environments, users may ask about any database without specifying which DB they are referring to, rendering table retrieval significantly more challenging (see Appendix~\ref{appendix:tbr_simulation} for more details on the simulation setup).}
In this simulation setup, the retrieval service does not achieve perfect recall at $k \geq 13$, which is the maximum number of tables per database in BIRD dev. \yeounoh{This is because tables are retrieved from all DBs (100 tables) without specifying the target DB. In Section~\ref{sec:experiment_all_tables} we explore the impact of including all tables for each DB and across DBs to ensure perfect recall at the cost of lower precision.}

\yeounoh{Here, we improve schema linking by including all database tables to guarantee that all relevant elements are provided along with lots of extra irrelevant ones. This raises concerns that the model might get confused or ``lost in the middle'' of a large context window  \citep{Floratou2024NL2SQLIA,liu-etal-2024-lost}. However, we observe that the latest long-context LLM, \textit{gemini-1.5}, does not exhibit this issue and shows ``near-perfect'' retrieval capability~\citep{reid2024gemini}.}
In Section~\ref{sec:experiment_all_tables} we compare our long-context schema linking approach to the baseline using top-K relevant tables, and in Section~\ref{sec:ablation} we study the increased context size and latency implications.
\\\\
\noindent\textbf{\textit{Column description and sample values  for improved column selection.}} Prior research \citep{wang-etal-2020-rat} shows that the column descriptors and values as part of prompt can improve the accuracy of schema linking and column selection. \citep{maamari2024death} used column descriptors, data types, and a few sample column values in the input prompt. These extra information is shown to help LLMs reason about the column references. In this work, we provide significantly more than just a few sample column values for any non-trivial `text` columns. In Section~\ref{sec:disambiguation_experiment}, we demonstrate how such a brute-force approach can fully leverage the model's ``near-perfect`` retrieval capability~\citep{reid2024gemini} to address ambiguous column reference and literal errors.
\\\\
 \noindent\textbf{\textit{User provided hints for additional clarification.}}
 The extended context window allows for additional instructions and clarification from user to be included in addition to the database schema and other column metadata.

\begin{figure}[!htbp]
\centering
\begin{tikzpicture}
  \node[draw, rectangle, fill=mycolor] (rect) {
    \begin{tabular}{@{}p{\linewidth}@{}} 
      \toprule
      \begin{minipage}{\linewidth}
        \footnotesize Question: What is the total number of \textbf{non-chartered} schools in the county of Los Angeles with \textbf{a percent (\%) of eligible free meals for grades 1 through 12 that is less than 0.18\%}?\\\\
        \footnotesize Evidence: non-chartered schools refer to schools whose \textbf{Charter = 0}; K-12 means grades 1 through 12; percent of eligible free rate for \textbf{K-12 = `Free Meal Count (K-12)` * 100 / `Enrollment (K-12)`}\\\\
        \footnotesize SQL: SELECT COUNT(T2.School) FROM frpm AS T1 INNER JOIN schools AS T2 ON T1.CDSCode = T2.CDSCode WHERE T2.County = 'Los Angeles' AND \textbf{T2.Charter = 0 AND CAST(T1.`Free Meal Count (K-12)` AS REAL) * 100 / T1.`Enrollment (K-12)` < 0.18}
    \end{minipage} \\
    \bottomrule
    \end{tabular}
  };
\end{tikzpicture}\vspace{-2mm}
\caption{A hint (Evidence) prescribing the nuanced column reference and the mathematical expression needed to answer a challenging question from BIRD dev.}
\vspace{-2mm}
\label{fig:hint}
\end{figure}
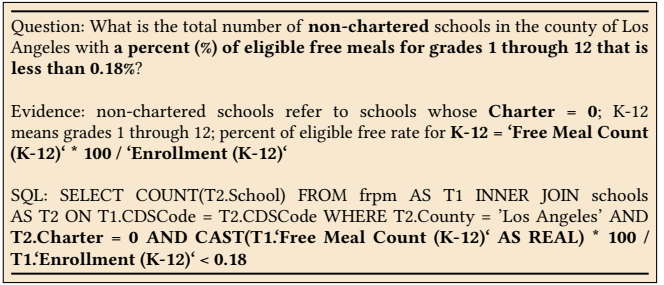

BIRD benchmark includes hints for accurate SQL generation, clarifying which column is required and/or how to compute the values. Fig.~\ref{fig:hint} shows an example question with hints. Not all BIRD dev questions have hints. In production NL2SQL services, the hints may come via  multi-turn interactions with the user. 
We provide the hints, if available, as part of the user question. 
In Section~\ref{sec:ablation} we discuss the impact of hints.
\\\\
\noindent\textbf{\textit{Synthetic examples for many-shot ICL:}}
While prior works on ICL for NL2SQL focused on selecting a few handful of examples (3-5) for ICL \citep{nan2023enhancing,pourreza2024din},  we generate many examples for many-shot ICL. Many-shot ICL is shown to improve LLM generation quality on many different domains 
\citep{agarwal2024manyshot}. Here we generate and use tens and hundreds of examples with the extended context window. The examples are also generated at inference time, as prescribed in \citep{pourreza2024chase}.
The online generation allows bypassing the example retrieval process and generate examples using schema elements more relevant to the user question (we pass the entire schema for a given database along with LLM-based relevant column selection results). The model generates synthetic examples including both the output SQL and input natural language question. The generated SQL queries follow the target benchmark datasets with respect to the common SQL features 
The structural similarities can help guide the model to generate with similar SQL structures and features.

    

In Section~\ref{sec:synthetic_example_selection} we demonstrate how synthetic examples based  many-shot ICL can improve the performance and  discuss the performance and cost trade-off of example generation in Section~\ref{sec:ablation}.
\\\\
\noindent\textbf{\textit{Relevant SQLite documentation sections:}}
The large extended context window allows passing SQLite documentation in big chunks, chapters and/or sections. Documentation writing tends to be lenghty and contains a few illustrative examples. 
We explore the feasibility and the impact of documentation on performance and cost in Section~\ref{sec:sqlite_documentation}.

\subsection{Fix \& Rewrite}\label{sec:fix_rewrite}
We allow the model to fix its errors and rewrite the SQL output based on execution. If the generated SQL query returns  an error during execution, it triggers correction modules \citep{wang2024macsql,caferouglu2024sql} \yeounoh{and retries until a valid SQL is returned within a fixed number of times}. \citep{pourreza2024din} showed that LLMs can correct minor syntax errors in SQL. While the existing self-correction mechanisms and the latest LLMs themselves are effective in fixing SQL syntax errors, they do not address more nuanced semantic errors like  invalid literal references and incorrect join paths. Semantic errors may persist after syntax error correction (e.g., referencing a valid literal value for a wrong column), and they are hard to detect without the ground-truth query results. We check for potential semantic errors if the query returns empty result; \yeounoh{we ask the model to rewrite such queries by providing more extensive list of sample column values. There is a risk of false positives in relying on a empty results for detecting semantic errors.  If the correct ground-truth result is empty, this could incur additional computational costs due to unnecessary retries. If the question is not ambiguous, the model should return the same correct empty result even when additional sample column values are included. However, if the question is ambiguous and the model re-generates a non-empty result, then we prefer that over empty result and move to the next verification step.} Seeing the full list of column values enables the model to select the right keyword and columns and to reason about more accurate alternative join paths. Passing the full lists of sample column values inflate the context size significantly and requires long context and LLMs that can retrieve accurately over the long context. We evaluate the impact of this long context disambiguation strategy in Section~\ref{sec:disambiguation_experiment}. This technique is expensive, but it can complement imperfect table \& column selection and schema linking, which often leads to sub-optimal NL2SQL performance~\citep{caferouglu2024sql}.

\subsection{Verify}
We use the untuned \textit{gemini-1.5-pro} LLM to verify the correctness of the final output. The model is given the entire database table schema and question (with additional hints, if provided) to judge. 
We believe that fine-tuning a verifier or a picker for self-consistency\citep{wang2022self,pourreza2024chase, 10.5555/3666122.3668142} can further improve accuracy. While we focus on leveraging the long context ICL for a typical NL2SQL pipeline with simpler techniques (self-correction, verification without fine-tuning), the aforementioned techniques are orthogonal and can be used in conjunction. We demonstrate how this verification step  can impact the final performance and the generation cost in Section~\ref{sec:ablation}.

\section{Detailed Analysis of Long Context NL2SQL Techniques }\label{sec:experiments}

\subsection{Evaluation setup}
We use the public Google Cloud (GCP) Vertex AI \textit{gemini} API for all our experiments. This is important for reproducibility of our experiments and analysis. We use the latest publicly available \textit{gemini-1.5-pro-002} and \textit{gemini-1.5-flash-002} checkpoints. \textit{gemini-1.5} is a long-context LLM class with up to 2-million token context for \textit{pro} and 1-million token context for \textit{flash}. Our test VMs and Vertex AI endpoints are located in the same GCP region (us-central1-b).

\yeounoh{
We report the full pipeline performance on various NL2SQL benchmark datasets (BIRD~\citep{li2024can}, SPIDER 1.0~\citep{Yu.al.18c}, KaggleDBQA~\citep{lee-2021-kaggle-dbqa} and BEAVER~\citep{chen2024beaver}) in Section~\ref{sec:benchmarks}. We run our micro-benchmark experiments using two widely used benchmark datasets, BIRD dev and KaggleDBQA test. BIRD is currently the most popular NL2SQL benchmark with a leaderboard (all the recent NL2SQL research publications use this dataset to compare their performance), and it contains the largest number of questions with varying degrees of difficulties spanning over multiple tables with varying schema complexity. KaggleDBQA is another benchmark dataset for Text-to-SQL tasks that focuses on real-world, cross-domain data sources.}

For the performance and cost metrics, we use popular execution accuracy (Ex Acc) and the accumulated tokens per request ($L$), and/or normalized latency per request ($T$) as the absolute latency measure behind the public API endpoints can vary from time to time -- a single \textit{gemini-1.5-pro} request latency with 8k-token is used as a reference unit ($T$=1.0); we report floating-point values to one decimal place, ignoring differences below 10\%.
\yeounoh{Likewise, we refrain from reporting the monetary cost directly, as it depends on the pricing model~\citep{gemini_pricing}, which is subject to change. Instead, we report the accumulated number of tokens per request $L$, which is strongly correlated with total monetary cost and remains invariant to pricing model changes.}

There is a strong positive correlation between $L$ and $T$, as discussed in Section~\ref{sec:experiment_latency}, and we report both in the ablation study. We report the average metrics across requests with low variance (margin of error/variability less than 2\%); if the observed variance is large, we report the mean plus one standard deviation (e.g., $\bar{L} + \sigma$) to bound the majority of requests.

Our ``full'' pipeline employs all the contextual information and the techniques as  discussed in Section~\ref{sec:approach} and Section~\ref{sec:ablation}; for the individual experiments, we use the ``baseline'' pipeline to study specific questions being addressed. The baseline pipeline includes entire DB schema, sample column values, instructions and hints, but excludes self-correction, disambiguation, synthetic examples and verification.

\subsection{Benchmark evaluation}\label{sec:benchmarks}

\begin{table}[th!]
\begin{minipage}{\columnwidth}
\centering
\caption{Benchmark performance and generation latency $T$ using the full long-context NL2SQL pipeline and with \textit{gemini-1.5} models. $T_g$ measures per request output generation time, including self-correction and retries.}\vspace{-2mm}
\label{tab:benchmarks}
\resizebox{\columnwidth}{!}{
\begin{tabular}{l |c c|c c }
\toprule
 & \multicolumn{2}{c|}{\textit{gemini-1.5-pro}} & \multicolumn{2}{c}{\textit{gemini-1.5-flash}}\\
 & Ex Acc (\%) & $\bar{T_{g}}$ (T units/req) & Ex Acc (\%) & $\bar{T_g}$ (T units/req) \\
\midrule
BIRD dev & 67.41 & 12.3  & 66.49 & 8.6 \\
SPIDER 1.0 test & 87.10 & 1.5 & 84.60 & 1.1 \\
\yeounoh{KaggleDBQA test} & 61.10 & 2.7 &58.40 & 1.8  \\
\yeounoh{BEAVER dw\footnote[1]{Only Data Warehouse 1 (dw) dataset with 48 questions is publicly available as of January  7th, 2025.}} & 60.41  & 506.1 & 50.00 & 506.1\\
\bottomrule
\end{tabular}
}
\end{minipage}
\vspace{-2mm}
\end{table}

\yeounoh{
Table~\ref{tab:benchmarks} shows the full pipeline evaluation results across various benchmark datasets. The popular benchmark datasets (BIRD dev, SPIDER 1.0 test, KaggleDBQA test) provide a diverse set of questions with varying difficulties across multiple domains, allowing us to compare our performance with other state-of-the-art approaches. 
Our approach leveraging the extended context of \textit{gemini-1.5} model can yield competitive results across all the benchmarks, without techniques like fine-tuning and self-consistency (Table~\ref{table:bird_dev_results}).
The newer benchmark for enterprise warehouse data (BEAVER dw) is particularly noteworthy because its business-oriented questions and real enterprise tables are significantly more complex and larger. Among the three widely used benchmarks, BIRD is the most complex, with questions averaging 0.918 JOINs over 6.82 tables per database. In contrast, BEAVER questions involve 4.25 JOINs over 105.0 tables per database, making it substantially more challenging. BEAVER is still in its early stages as a benchmark. The \textit{dw} dataset we used had some issues, like UNIQUE constraint violations on primary keys for which we ignored and allowed duplicates. Our long context pipeline achieves Ex Acc of 60.41\% (48.64\% ignoring questions with null golden queries) with \textit{gemini-1.5-pro} without using the golden tables, but the entire DB tables and the column mappings, similar to the hints in BIRD benchmark. The The best 1-shot Ex Acc on BEAVER (dw + 45 additional questions), reported by \citep{chen2024beaver}, is 2.0 in PL/SQL using \textit{gpt-4o} with relevant table schema retrieval, and 0.0 without retrieval. 
}

\yeounoh{
The normalized generation latency $T_g$ increases with both larger context size and task difficulty (question and schema complexity), as it includes self-correction and retries (using exponential backoff, with a maximum of 5 retries per request). For instance, BEAVER takes significantly longer to generate on average because our long context pipeline produces context size exceeding the limits (2-million tokens for \textit{pro} and 1-million tokens for \textit{flash}) or results in semantically invalid (empty) outputs, in which case the model retries a fixed number of times until it finds a better answer. Consequently, for complex large datasets, like BEAVER, \textit{flash} can paradoxically take as long as or even longer than \textit{pro} with more retries due to the smaller context size limit.
}

In general, \textit{gemini-1.5-pro} is better at complex reasoning tasks like NL2SQL, whereas \textit{gemini-1.5-flash} is more cost-effective (lower pricing per token). 
We have noticed that the latest releases of the \textit{gemini} models (released for public access on Sep. 24\textsuperscript{th}, 2024) show significant performance improvements, and most notably, \textit{flash} now performs more comparably to \textit{pro} on the select benchmarks. This makes the more cost-efficient \textit{flash} a highly attractive choice for production NL2SQL.

\subsection{Schema linking with all database tables}
\label{sec:experiment_all_tables}

We explore the impact of passing the entire table schema set collected from entire databases. Schema linking is such a critical component that we strive to achieve a high-recall in production. \yeounoh{With a long-context model, we can include a significantly larger set of table definitions and schemas instead of limiting the input to only the most relevant top-K tables.
As is standard practice, we retrieve the most relevant tables based on the NL question and the table DDL statement, using embedding similarity between the two.  The top-k table retrieval is done across all the tables and DBs, mirroring the production setup illustrated in Figure~\ref{fig:tbr_simulation}. Alternatively, we provide all the tables from a given target DB (benchmark datasets specify to which DB the question is referring to) or from all the tables in the dataset (across all DBs), without the top-k retrieval.}
The goal is to ensure near perfect (higher) recall at the cost of lower precision \yeounoh{by providing more tables, if not all, via long context}. 

\begin{table}[t!]
\centering
\caption{Table retrieval (TBR) performance and execution accuracy with the baseline pipeline. Putting all tables (\textit{all tables / DB} or \textit{all tables / dataset})  
 guarantees perfect recall}
\vspace{-3.5mm}
\label{tab:tbr_bird_ex_acc}
\subcaption{BIRD dev}
\vspace{-2mm}
\resizebox{\columnwidth}{!}{
\begin{tabular}{l c c c c}
\toprule
\# tables (k) & k=1 & k=7 & all tables / DB  & all  tables / dataset\\
\midrule
Ex Acc (\%) & 38.01 & 54.69 & 62.32 & 62.58 \\
TBR Recall (\%)& 45 & 82 & 100 & 100 \\
TBR Precision (\%) & 77 & 23 & Low ($< 35\%$) & Low ($< 2\%$)  \\
\midrule
$\bar{L}$ (tok/req) & 2002.54 & 4627.68 & 7380.86 & 72619.96 \\
\bottomrule
\end{tabular}
}

\smallskip
\subcaption{\yeounoh{KaggleDBQA test}}
\vspace{-2mm}
\resizebox{\columnwidth}{!}{
\begin{tabular}{l c c c c}
\toprule
\# tables (k) & k=1 & k=2 & all tables / DB  & all  tables / dataset \\
\midrule
Ex Acc (\%) & 43.24 & 50.27 & 56.75 & 60.54 \\
TBR Recall (\%)& 81.53 & 91.26 & 100 & 100 \\
TBR Precision (\%) & 90.27 & 53.51 & 68.86 & 7.44  \\
\midrule
$\bar{L}$ (tok/req) & 922.29 & 1731.64 & 1485.45 & 26087.86 \\
\bottomrule
\end{tabular}
}\vspace{-3mm}
\end{table}

The results shown in Table~\ref{tab:tbr_bird_ex_acc}  \yeounoh{demonstrate that the model does not get confused despite the presence of a large number of mostly irrelevant table definitions in the context. Thus, providing more tables, rather than solely relying on an imperfect table retrieval mechanism, can be beneficial—provided the context size allows for it. 
The result (for $k=1$) also re-affirms that} schema linking is critical, since without all the ground truth tables in the prompt ($k=1$) the model cannot generate good quality SQL outputs. 
\yeounoh{Conversely, providing all tables from the target DB} ensures perfect TBR recall and results in higher Ex Acc \yeounoh{in both BIRD dev and KaggleDBQA test}. This means that TBR recall is more important than precision or accuracy as the model generation quality does not degrade much with many irrelevant tables (68.18 irrelevant tables on the average for BIRD dev \yeounoh{and 14.875 for KaggleDBQA). The \textit{all tables / DB} setting utilizes up to 13 tables per request (5 tables for KaggleDBQA), depending on the target database in BIRD dev; the average number of tables across different databases is 6.82 (2.125 for KaggleDBQA)}.
\yeounoh{It is interesting to note that the execution accuracy result for \textit{all tables / dataset} is slightly higher than that of \textit{all tables /DB}. The difference is more pronounced in the KaggleDBQA test (almost +3.8\% Ex Acc), whereas it falls within the 5\% margin of error for BIRD dev (we use medium temperature for LLM generation, and Ex Acc varies slightly per run).}

\yeounoh{This suggests that the model's output quality is at least comparable, and the inclusion of additional irrelevant schema information does not degrade generation quality.} 

\yeounoh{
On the one hand, including as many tables as possible (e.g., user history, all other tables within the target database) ensures high TBR recall, which long-context LLMs can accommodate. On the other hand, the latency cost increases with larger context sizes (see Fig.~\ref{fig:latency_experiment}), and beyond a certain point, adding more irrelevant tables does not provide additional benefits to justify the increased cost.  Furthermore, some production settings require processing hundreds of very wide tables, making it prohibitively expensive to include all tables in the context (see Section~\ref{sec:production_limitation} for more discussion). Thus, accurate retrieval and filtering remains desirable, and leveraging long-context models can serve as a compensatory mechanism to offset imperfect retrieval for higher recall.
}

\vspace{-2mm}
\subsection{Many-shot ICL with example selection}
Here, we evaluate the impact of many-shot in-context learning (ICL) using examples selected from BIRD train dataset. The examples are retrieved based on the question embedding similarity.  

\begin{table}[t!]
\centering
\caption{Ex Acc (\%) with the baseline pipeline \& selected similar examples from the train datasets, $\sigma$(train), and ground-truth (+ GT)}\vspace{-2mm}
\label{tab:many_shot_bird_train}
\subcaption{Varying number of similar train examples and GT inserted in the middle}
\resizebox{\columnwidth}{!}{
\begin{tabular}{l l c c c c c}
\toprule
& \# train examples & 0 & 5 & 20 & 50 & 100 \\
\midrule
\multirow{3}{*}{BIRD dev} & $\sigma$(train) & 61.60 & 63.17 & 62.71 & 62.58 & 62.52 \\
& $\sigma$(train) + GT & 78.68 & 77.18 & 77.84 & 77.51 & 78.81 \\
\cline{2-7}
& $\bar{L}$ (tok/req) & 7380.86 &  8000.67 & 9927.24 & 13808.71 & 20358.30 \\
\midrule
\multirow{3}{*}{\begin{tabular}[c]{@{}l@{}} KaggleDBQA \\ \yeounoh{test} \end{tabular}} & $\sigma$(train) & 57.83 & 65.40 & 79.45 & 64.32 & 65.94 \\
& $\sigma$(train) + GT & 80.50 & 81.62 & 78.91 & 81.08 & 80.50 \\
\cline{2-7}
& $\bar{L}$ (tok/req) & 922.29 &  1753.64 & 2550.10 & 4191.18 & 6266.45 \\
\bottomrule
\end{tabular}
}
\centering
\vspace{3mm}
\subcaption{GT inserted at different  positions (normalized) within the context window with selected 100 $\sigma$(train) + GT}
\resizebox{0.8\columnwidth}{!}{
\begin{tabular}{l c c c c c}
\toprule
GT position & 0.1 & 0.25 & 0.50 & 0.75 & 0.9 \\
\midrule
BIRD dev   & 77.77 & 78.29 & 78.1 & 78.42 & 78.16 \\
KaggleDBQA test   & 81.62 & 81.62 &  80.00 & 80.00 & 80.08 \\
\bottomrule
\end{tabular}
}
\vspace{3mm}
\subcaption{\yeounoh{Changing the order of the examples block (100 $\sigma$(train) + GT) in the context: beginning (before the instructions), middle (before the schema details), end (after the schema details)}}
\resizebox{0.7\columnwidth}{!}{
\begin{tabular}{l c c c}
\toprule
Block position & beginning & middle & end \\
\midrule
BIRD dev  & 80.44 & 78.23 & 78.29 \\
KaggleDBQA test &78.91 & 79.45 & 80.00 \\ 
\bottomrule
\end{tabular}
}\vspace{-3mm}
\end{table}

Table~\ref{tab:many_shot_bird_train}(a) shows that the train examples did not provide much boost and instead resulted in worse performance. This means that the model’s ability to learn from similar examples from BIRD train is quite limited and it can be even distracted by them. Also, notice that the impact of distraction seems to be limited, say going from 1 similar example (11k tokens) to 100 similar examples (50k tokens).
To further examine the model’s sensitivity to many irrelevant train examples, we injected 1 ground-truth example from BIRD dev in the “middle” of the example blob (the average context size is omitted as it should be almost identical to the previous table). 
While the model can select the ground truth example among many other irrelevant ones, the presence of those  irrelevant train examples can confuse the model as opposed to just having that ground truth. The recall matters more than precision, but the model is still sensitive to bad examples (low precision) to a limited degree -- note that the accuracy with 100 random examples and 1 ground truth is much higher than no examples at all. 
It is also important to note that the model is robust to both the position and ordering of the ground-truth example.
\yeounoh{\citep{liu-etal-2024-lost} showed that LLMs tend to emphasize information at the beginning and end of the context window, highlighting the importance of proper ranking and ordering for retrieval-augmented generation (RAG).
As in Table~\ref{tab:many_shot_bird_train}(b-c), where the GT is injected at different positions within the context window,} we observe that with \textit{gemini-pro-1.5} the ordering \yeounoh{of the examples (and the schema information relocated by the relocation of the examples in (c))} has a negligible impact, and the long-context LLMs can retrieve the relevant ground truth example \yeounoh{(also the schema information)} from anywhere within its context window.

The model’s ability to learn and generalize from random examples is limited: a single relevant example is far better than having many random examples. While it is crucial to capture the relevant examples in the context (relevant examples boost the generation quality), it is also very challenging to capture the relevant ones based on the user questions. In the next section, we also share our experience generating many synthetic examples per question to skip the example selection step.

\subsection{Synthetic example generation VS. example selection}
\label{sec:synthetic_example_selection}

\begin{figure}[ht!]
    \centering
    \begin{subfigure}{0.495\columnwidth} 
        \centering
        \centering
        \includegraphics[width=\columnwidth, height=0.7\columnwidth]{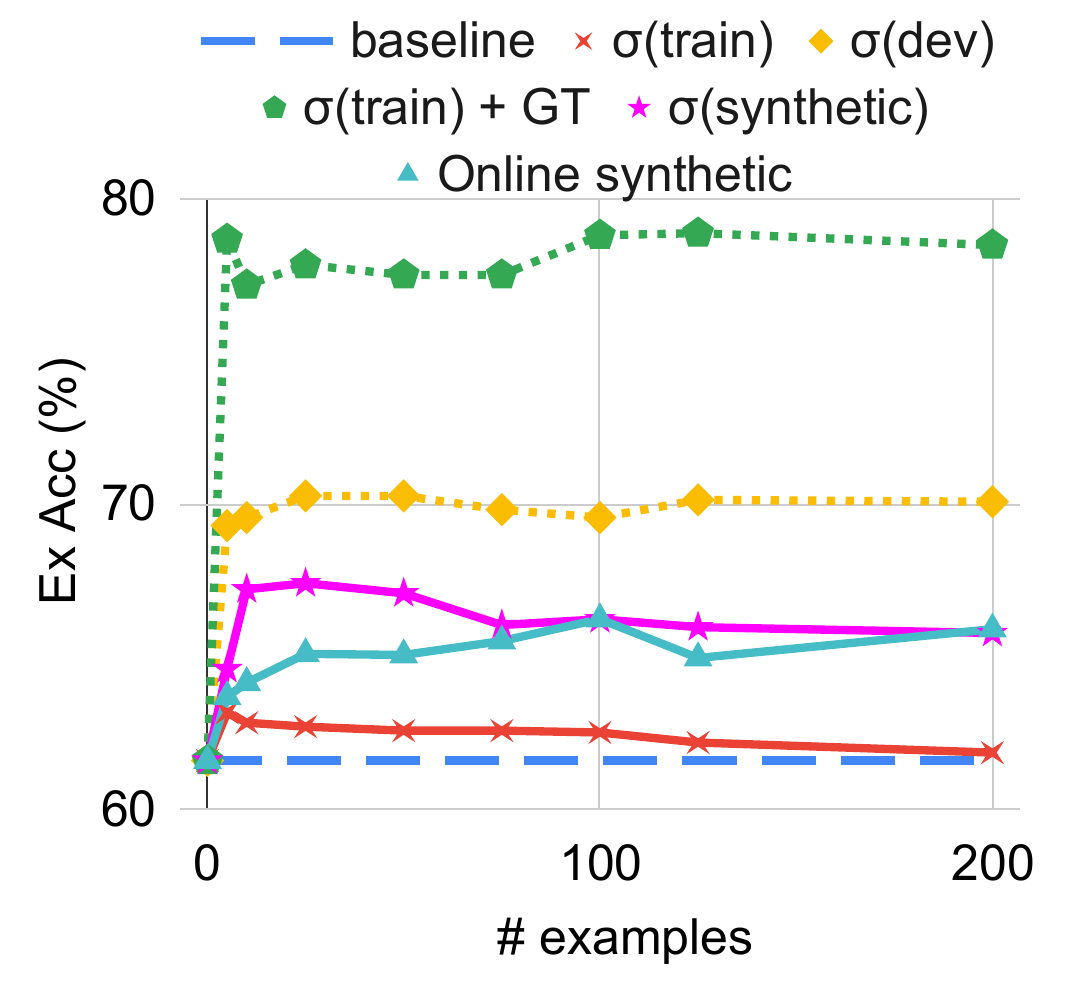} 
        \caption{BIRD dev}
        \label{fig:subfig1}
    \end{subfigure}
    \hfill 
    \begin{subfigure}{0.495\columnwidth} 
        \centering
        \includegraphics[width=\columnwidth, height=0.7\columnwidth]{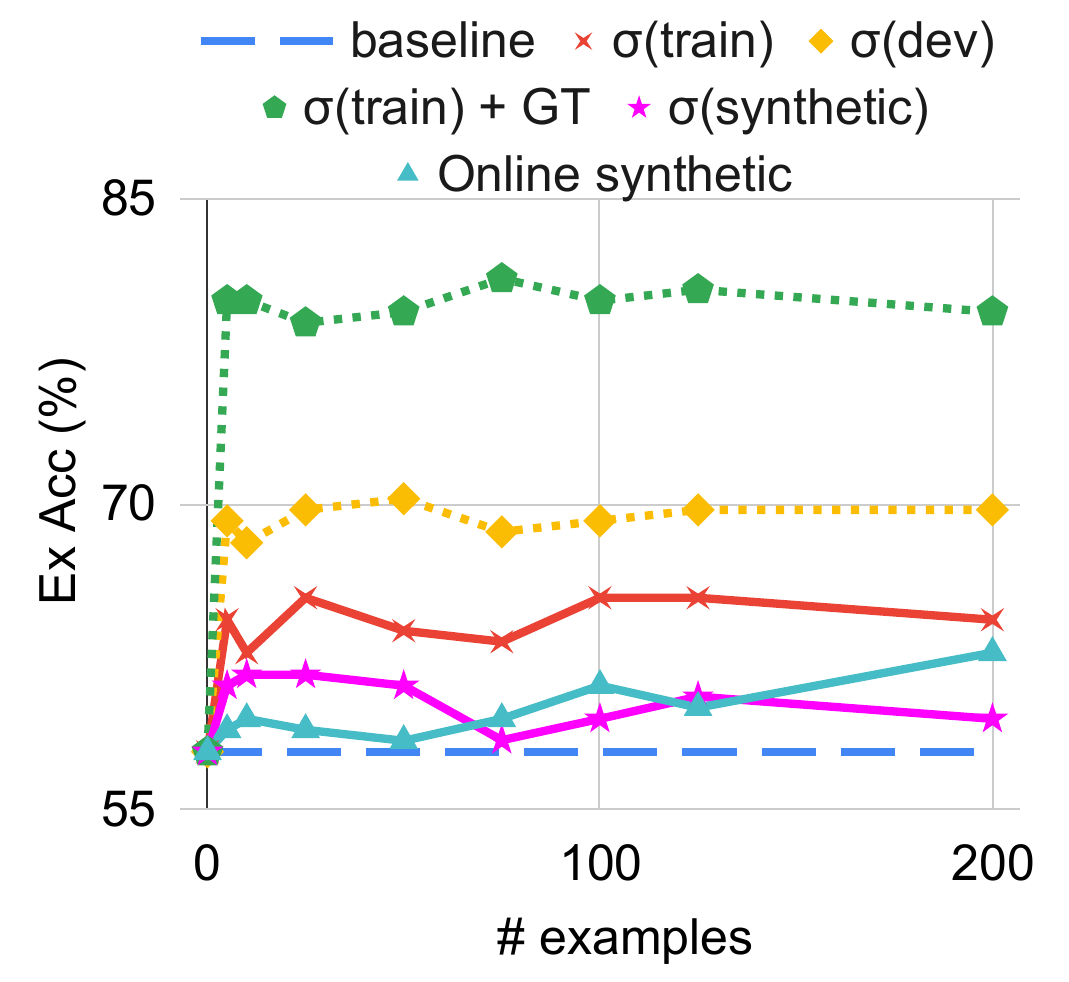} 
        \caption{\yeounoh{KaggleDBQA test}}
        \label{fig:subfig2}
    \end{subfigure}
    \caption{Question similarity-based example selection ($\sigma$) vs. online example generation. The baseline uses no examples; examples were retrieved from BIRD \textit{train} dataset, \textit{dev} dataset (excluding the GT) and \textit{synthetic} generated examples. \textit{train + GT} uses the retrieved \textit{train} examples and the ground-truth SQL from \textit{dev}.}
    \label{fig:example_selection}\vspace{-2mm}
\end{figure}

Fig.~\ref{fig:example_selection} illustrates \yeounoh{how synthetic examples improve the quality of generated SQL query and compares with other types of examples retrieved from different example pools. As is a standard practice, the example retrieval was done via question embedding similarity using \textit{gecko}\footnote{https://cloud.google.com/vertex-ai/generative-ai/docs/embeddings/get-text-embeddings} embedding model. 
Note that using the ground-truth SQL queries from \textit{\textit{dev}} dataset provides  performance ceilings ($\sigma$(\textit{dev}) and $\sigma$(\textit{train} + GT)), but is not allowed without an oracle (i.e., using the ground truth queries from the evaluation dataset). While \textit{dev} excludes the specific GT query for a given question, it can still retrieve similar examples directly from the ground truth DB in the dev dataset. The gap between $\sigma$(\textit{dev}) and $\sigma$(\textit{train} + GT) indicates that providing the GT as one of the many examples is better than retrieving dev examples that are similar to the GT. 
$\sigma$(\textit{train}) is the common practice where the relevant examples are selected from available datasets, such as the training dataset (\textit{train}). In the case of BIRD dev (a), providing relevant \textit{train} examples does as least as good as the baseline (no example) with a slight improvement when fewer than 200 shots are provided. However, \textit{synthetic} examples ($\sigma$(\textit{synthetic}) and \textit{online synthetic}) provide significant uplift over $\sigma$(\textit{train}). The gap between $\sigma$(\textit{synthetic}) and \textit{original synthetic} indicates that one can be more economical with the context size (achieving higher performance with a fewer examples) by filtering the relevant \textit{synthetic} examples. In case of KaggleDBQA test (b), both training and synthetic examples boost the Ex Acc, while \textit{train} performs much better than the synthetic ones. This is because BIRD train and dev datasets are sampled from different database domains, whereas KaggleDBQA train and test datasets are sampled from the same domains.  It is important to note that the many-shot ICL using synthetic examples can help further improve Ex Acc, and even more so in the cross-domain case (e.g., BIRD dev) where the available training examples are less relevant.
}

\subsection{Self-correction with entire schema and more column values}\label{sec:disambiguation_experiment}
\begin{table}[t!]
\centering
\begin{minipage}{\columnwidth}
\caption{Ex Acc (\%) and extra accumulated tokens for correction. The eval baseline pipeline does not include self-correction (SC); \textit{disambiguation} refers to providing extensive sample column values as described in Section~\ref{sec:fix_rewrite}; \textit{filtered schema} is providing filtered relevant columns and values to re-write}
\label{tab:self_correction}
\resizebox{\columnwidth}{!}{
\begin{tabular}{l c c c}
\toprule
& \textbf{Corrected SQL} & \multicolumn{2}{c}{\textbf{Tokens used for Correction}}\\
& Ex Acc (\%) & $\bar{L}$  (tok/req)& $\bar{L} + \sigma$  (tok/req)\\
\midrule
 \multicolumn{4}{c}{BIRD dev} \\
\midrule
Baseline (w/o SC) & 61.6 & - & -  \\
SC & 64.80 & 3632.77 & 6631.60  \\
SC + disambiguation & 65.51 & 15754.64 & 334999.28  \\
SC + filtered schema & 65.84 & 3930.55\textsuperscript{a}  & 7001.89 \\
\midrule
\multicolumn{4}{c}{\yeounoh{KaggleDBQA test}} \\
\midrule
Baseline (w/o SC) & 58.91 & - & -  \\
SC & 59.45 & 2211.34 & 34573.67  \\
SC + disambiguation & 61.08 & 111238.44 & 205852.86  \\
SC + filtered schema & 59.45 & 2335.9\footnote[1]{this excludes the cost of running the LLM-based column selection, where multiple LLM calls are needed to extract relevant columns for given request}  & 35475.69 \\
\bottomrule
\end{tabular}
}
\end{minipage} \vspace{-5mm}
\end{table}

Table~\ref{tab:self_correction} shows how LLM's self-correction can help with SQL generation and its refinement. 
We compare LLM-based self-correction (SC) with two more advanced techniques: \textit{disambiguation} and \textit{filtered schema}. \textit{disambiguation} is the long-context enabled technique where an extended lists of sample column values are shown to the model upon detecting empty results from syntactically correct SQL; \yeounoh{\textit{filtered schema} uses the column selection results to filter relevant schema elements per user question, following the approach of \citep{talaei2024chess}}. Both \textit{disambiguation} and \textit{filtered schema} can help LLMs correct for any invalid literal and column references. \yeounoh{One potential issue with \textit{disambiguation} is that the extra column values can distract the model, especially in the case of false positive detection (e.g., when the correct answer is null). However, empirical results show that it generally improves performance, performing slightly better or at least as well as \textit{SC}.}
While comparable in performance, the techniques vary in terms of the cost. The accumulated average number of prompt tokens used for correction increases significantly for \textit{disambiguation}, as it includes full schema and extra sample column values.  

Accurate column selection is crucial for improving NL2SQL performance, and \textit{filtered schema} does not increase the context size during correction. However, preparing \textit{filtered schema} requires an additional retrieval step, which incurs a cost (see  Appendix~\ref{appendix:impact_col_selection} for more discussion).
\textit{filtered schema} is a great strategy, if accurate column selection process is available. 
However, in our final long-context pipeline, we use SC +  \textit{disambiguation} to avoid extra retrieval/selection step and do everything in-context.
Due to the high variance in token usage, we report the mean tokens per request plus one standard deviation (1-S.D) to capture the range used for most requests (~68\%).

\subsection{In-context learning with SQLite documentation}\label{sec:sqlite_documentation}
We test whether the model can effectively learn from SQLite documentation and improve its SQL generation quality. We downloaded the entire SQLite documentation from its official homepage~\citep{sqlite_home_page_2024}. The original documentation comprises 16 million tokens across 792 HTML files, each covering a distinct theme. We applied two strategies to split the entire documentation into chunks so that we can augment the prompts with relevant information for ICL. The coarse-grained chucking strategy split the documentation by HTML file, while the fine-grained chucking strategy further separate each file by sections which ends up with 4125 smaller chunks in total. Similar to example retrieval, we embed the natural language questions and document chunks into vectors with the Gecko text embedding model~\citep{lee2024gecko} and employ nearest neighbor search to identify the most relevant document chunks to the given question. We could further compress the context via summarization, but decided not to, since it drops important details, like syntax rules and examples. 

\begin{table}[t!] 
\centering
\caption{Ex Acc with SQLite  document chunks, measured with the baseline pipeline.
}\vspace{-2mm}
\label{tab:sqlite_document_icl}
\subcaption{Coarse-grained document chunks}
\vspace{-2mm}
\resizebox{\columnwidth}{!}{
\begin{tabular}{l l c c c c}
\toprule
& \# chunks & 0 & 1 & 2 & 3 \\
\midrule
\multirow{2}{*}{BIRD dev}& Ex Acc (\%)  & 61.84 & 61.54 & 61.08 & 61.67 \\
& $\bar{L}$ (tok/req) & 7380.87 & 32703.21 & 42087.60 & 51308.69 \\
\midrule
\multirow{2}{*}{\begin{tabular}[c]{@{}l@{}} KaggleDBQA \\ \yeounoh{test} \end{tabular}}& Ex Acc (\%)  & 57.83 & 57.83 & 59.45& 60.00\\
& $\bar{L}$ (tok/req) & 922.29 & 5224.18 & 11441.86 & 17209.37 \\
\bottomrule
\end{tabular}
}

\smallskip

\subcaption{Fine-grained document chunks}
\vspace{-2mm}
\resizebox{\columnwidth}{!}{
\begin{tabular}{l l c c c c c}
\toprule
& \# chunks & 0 & 1 & 5 & 10 & 15 \\
\midrule
\multirow{2}{*}{BIRD dev} & Ex Acc (\%)  & 61.84 & 60.43 & 61.86 & 61.54 & 61.15 \\
& $\bar{L}$ (tok/req) & 7380.87 & 7610.66 & 10115.78 & 14244.70 & 17560.80 \\
\midrule
\multirow{2}{*}{\begin{tabular}[c]{@{}l@{}} KaggleDBQA \\ \yeounoh{test} \end{tabular}}& Ex Acc (\%)  & 57.83 & 59.45 & 57.29 & 58.37 & 57.29 \\
& $\bar{L}$ (tok/req) & 922.29 & 2187.10 & 6202.0 & 7240.81 & 9545.82 \\
\bottomrule
\end{tabular}
}
\vspace{-2mm}
\end{table}

We show the execution accuracy and context size for document retrieval in table~\ref{tab:sqlite_document_icl}. The retrieval is challenging because the natural language question does not reveal the full structure and features of the corresponding SQL query; furthermore writing a correct SQL query requires a combination of concepts from multiple sections of the documentation.  However, we believe that the imperfect retrieval is not the reason why the documentation adds little to no value to generation. It is actually the nature of the errors that the LLMs make, which are mostly semantic errors (e.g., executable queries returning semantically irrelevant results).

The \textit{Gemini-1.5-Pro} model is already well-versed with the SQLite syntax and function details that are illustrated in the documentation. The SQLite documentation is likely already in the pre-train mixture. Furthermore, the model can self-correct itself for syntactic errors based on the error message and without any documentation or examples. A recent work \citep{talaei2024chess} based on GPT-4 also points out that there is no obvious errors based on simple syntax misuse (e.g., function names), but rather more subtle formatting (e.g., does not follow required date format) and semantic errors (output SQL is executable but incorrect w.r.t. the request). Reading the SQLite documentation would not help with such semantic errors – because the model is already producing executable queries that are semantically incorrect. While there is no accuracy gain, putting the documentation chunks increase the context size, thus the latency, significantly.

\subsection{Long context and latency relationship}
\label{sec:experiment_latency}

\begin{table}[t!] 
\centering
\caption{Average normalized generation and verification latency on BIRD dev using the baseline pipeline with \textit{gemini-1.5-pro} and \textit{gemini-1.5-flash}. The normalized latency time unit is set to the average generation latency on BIRD dev.
}
\vspace{-2mm}
\label{tab:latency_single_call}
\resizebox{\columnwidth}{!}{
\begin{tabular}{l c c}
\toprule
& gemini-1.5-pro & gemini-1.5-flash \\
\midrule
Single generation latency (units/req)  & 1.0 & 0.8 \\
Single verification latency (units/req) & 0.9 & 0.2 \\
\bottomrule
\end{tabular}
}\vspace{-2mm}
\end{table}

\begin{figure}[t!]
    \centering
        \centering
        \includegraphics[width=0.5\textwidth,height=0.5\columnwidth]{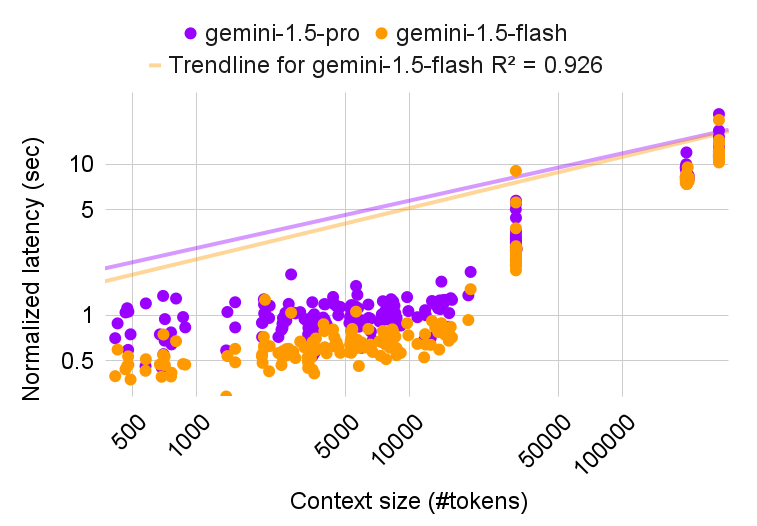} 
    \vspace{-5mm}
    \caption{Single LLM request latency by different context sizes. The axes are in log-scale, and there is a step-wise linear relationship between latency and context size. Both \textit{pro} and \textit{flash} models suffer from increased and high-variance latency per request beyond $>32k$ tokens context size.}
    \label{fig:latency_experiment}
\vspace{-0.4cm}
\end{figure}

In this section, we look at the context size (tokens) and latency relationship. \yeounoh{The relationship between context size and latency in Google's \textit{gemini-1.5} generation API is primarily a function of the number of tokens and is agnostic to the benchmark datasets. We evaluate this relationship using sampled requests of varying context size from BIRD dev.} Figure~\ref{fig:latency_experiment} illustrates that there is near-linear relationship (strong positive correlation $R^2$=92.6). It is interesting to note that the latency and the variance starts increasing significantly beyond context size $>$32k. The larger context LLMs require inference computation scattered across multiple hosts and accelerators, introducing additional queuing delays. While we expect the smaller \textit{gemini-1.5-flash} variant to exhibit lower latency, it also suffers from increased latency in the long tail due to queuing delays and differences in resource allocation. The generation latency increases significantly ($>>$ 4 seconds) with larger context size ($>$32k tokens), thus we should increase the context only if there is a clear gain in generation quality. Fortunately, the average relationship between context size and latency remains linear, making it easy to model.

Table~\ref{tab:latency_single_call} illustrates the average latency difference for the two \textit{gemini-1.5} model variants: the larger, more expensive \textit{pro} and the small, more cost-efficient \textit{flash}. \yeounoh{The average generation latency for BIRD dev is used as the unit for normalized latency $T$.} The difference in verification latency is more pronounced because the context size is smaller, whereas the generation with larger context \textit{pro} and \textit{flash} requires more prefill computation, which results in increased and similar queuing delay. Both models experience increased and similar average generation latency. For the single verification latency (indicative of the smaller context latency) \textit{flash} is almost 75\% faster than \textit{pro}.

\section{Ablation Study}\label{sec:ablation}
\begin{table*}[!ht]
\centering
\begin{minipage}{\textwidth}
\caption{Ablation analysis of the long-context NL2SQL pipeline. The pipeline shows strong performance on BIRD dev, even without any fine-tuning and/or multiple candidate generation (a.k.a., self-consistency). The context size per request also increases significantly, and we measure the total accumulated tokens over any  refinement iterations (self-correction) per given user request. We report 1-standard-deviation from the mean stats to bound majority of the requests.}
\label{tab:ablation_study}
\resizebox{\textwidth}{!}{
\begin{tabular}{l |c c c c  | c  | c }
\toprule
& \multicolumn{4}{c|}{\textbf{BIRD Dev Ex Acc (\%)}} & \multicolumn{1}{c}{\textbf{Context Size (tok/req)}} & \multicolumn{1}{c}{\textbf{Latency ($T$ units/req)}}
\\
Long Context NL2SQL & Simple & Moderate & Challenging & Overall & $\bar{L} + \sigma$ & $\bar{T} + \sigma$
\\
\midrule

+ All DB table schema & 52.22 ( - ) & 30.82 ( - )& 32.41 ( - )& 43.87 ( - ) & 10772.77 ( - ) & 1.0 ( - )\\
+ Hints & \cellcolor{green!25}67.35 ( \textbf{$\uparrow$ 15.13} )& \cellcolor{green!35}51.08 ( \textbf{$\uparrow$ 20.26} )& \cellcolor{green!25}44.14 (\textbf{$\uparrow$ 11.73})& \cellcolor{green!25}60.23  (\textbf{$\uparrow$ 16.36}) & 10796.12 ($\uparrow$ 23.35) & 1.0 ( - )\\
+ Sample column values & 68.11 ( $\uparrow$ 0.76)& 53.66 ( \textbf{$\uparrow$ 2.58})&\cellcolor{green!18} 49.66 ( \textbf{$\uparrow$ 5.52} )& 61.99 ($\uparrow$ 1.76) & 15568.47 ( $\uparrow$ 4772.35) & 1.0 ( - )\\
+ Self-correction & \cellcolor{green!10}71.03 ( \textbf{$\uparrow$ 2.92})& \cellcolor{green!10}56.25 ( \textbf{$\uparrow$ 2.59} )& \cellcolor{green!10}52.41 (\textbf{$\uparrow$ 2.75})& \cellcolor{green!10}64.80 ( \textbf{$\uparrow$ 2.81}) &\cellcolor{green!18} 50142.14 ( \textbf{$\uparrow$ 34573.67}) & 2.5 ($\uparrow$ 1.5)\\
+ Disambiguation & 71.14 ( $\uparrow$ 0.11)& 57.54 ( $\uparrow$ 1.29 )& \cellcolor{green!10}55.17 ( \textbf{$\uparrow$ 2.76} ) & 65.51 ( $\uparrow$ 0.71) &\cellcolor{green!35} 385141.42 ( \textbf{$\uparrow$ 334999.28}) & \cellcolor{green!18}33.5 (\textbf{$\uparrow$ 32})\\
+ Synthetic examples & 72.32 ( $\uparrow$ 1.18)& 59.05 ( $\uparrow$ 1.51 ) &\cellcolor{red!10} 53.79 (\textbf{$\downarrow$ 1.38 } )& 66.56 ($\uparrow$ 1.05) &
390413.05
($\uparrow$ 5271.62902) & \cellcolor{green!35}35.8 + 90.8\footnote[1]{We measure online example generation latency separately, which can be hidden/reduced  if pre-generated and retrieved instead.} (\textbf{$\uparrow$ 93.1})\\
\midrule
+ Verify \& retry\footnote[2]{The average number of attempts is 1.01, where the first output is accepted most of the time, except a few challenging cases.} & 72.65 ( $\uparrow$ 0.33) & 59.70 ( $\uparrow$ 0.65 ) & \cellcolor{green!15}55.86 ( \textbf{$\uparrow$ 2.07 } ) & 67.14  ($\uparrow$ 0.58 )& 399229.81 ($\uparrow$ 8816.75) & 37.1 + 90.8 ($\uparrow$ 1.3) \\
\bottomrule
\end{tabular}
}
\end{minipage}\vspace{-2mm}
\end{table*}

Overall, we focus on identifying useful information for improving NL2SQL performance -- and measure the impacts of the contextual information in terms of both performance and cost. We exclude the rules and SQLite documentation chunks from the ablation study, since their separate contributions are negligible. The rules are fixed and included as part of the instructions.
\yeounoh{We also omit certain popular techniques, such as Chain-of-Thought (CoT) prompting and self-consistency. CoT aids LLMs by breaking down complex tasks into subtasks, enabling multi-step reasoning for structured problems like NL2SQL~\citep{liu2023divide,tai2023exploring}. However, our observations indicate that CoT does not improve final output quality while significantly increasing context size—often by several thousand tokens.}

Table~\ref{tab:ablation_study} reports the generation performance (Ex Acc) along with accumulated context size (tokens/req) and latency (sec/req) per user request \yeounoh{using BIRD dev -- the most widely used NL2SQL benchmark, containing diverse questions and tables spanning multiple domains. We also report the results of a mini-ablation study using other benchmark datasets in Table~\ref{tab:mini_ablation}.
The study follows an incremental approach, adding one technique or contextual information element at a time.} The techniques and information that are more commonly used for NL2SQL are added earlier  (schema, hints, column samples and self-correction, respectively) and our long-context pipeline specific components are added later  (disambiguation and synthetic examples).

Because the latest long-context LLM, \textit{gemini-pro-1.5}, has strong retrieval capabilities, we observe overall performance improvements as more contextual information is included. However, when we categorize BIRD dev questions by difficulty level, we find that different types of contextual information benefit different types of queries. For instance, providing extra sample column values (or providing them all for disambiguation) is more helpful for challenging questions, where the question is more nuanced and/or ambiguous - allowing the model to leverage exact column values to choose the correct column and join paths. Synthetic examples show the opposite where it helps with simple and moderate questions, but hurt the challenging questions. Synthetic examples are generated to illustrate common SQL features and clauses using the schema elements from the target DB~\citep{pourreza2024chase}. They tend to be simpler and involve less ambiguous mappings between natural language questions and SQL queries. The hints are also interesting in that it appears to be one of the most critical ingredients for accurate SQL generation and it benefits moderate questions the most. BIRD datasets contain hints as to clarify nuanced concepts (e.g., ``eligible free rate for K-12'') with a correct expression (``Eligible free rate for K-12 = `Free Meal Count (K-12)` / `Enrollment (K-12)`''), and moderate queries often consist of more nuanced concepts that require such clarifications.

One key aspect of leveraging long context is cost. 
Table~\ref{tab:ablation_study} illustrates how each long context information or technique contributes to the overall context size and the latency, which are highly correlated (we discuss the near-linear relationship between context size and latency in Section~\ref{sec:experiment_latency}). For both context size $L$ and latency $T$, we report the average plus 1-S.D. because variance is high due to retries and self-correction mechanisms. This way, the reported measures capture the majority of the evaluated requests. 
Note that $L$ and $T$ track accumulated tokens and latency per request, until the final output is returned. Self-correction (retry) can be triggered for various reasons, ranging from simple syntax errors to empty results. The pipeline retries up to 5 times, applying appropriate fixes/modifications and increasing the temperature.
Due to its recursive nature, self-correction increases the latency (and the accumulated tokens) significantly; the cost grows even more exponentially with disambiguation (which involves including additional column value examples). The online synthetic example generation also adds to the runtime latency significantly, since the generation process involves a long sequence of auto-regressive decoding. However, if synthetic examples are generated offline and retrieved at runtime, as tested in Section~\ref{sec:synthetic_example_selection}, the generation latency can be reduced.
It is interesting to note that adding a few thousands tokens does not increase the latency that significantly over the entire pipeline. For instance, increase of 8816 tokens per request for ``verify \& retry'' delayed the latency only by 1.3 normalized latency units (mean + 1-S.D.).

\begin{table}[t!]
\centering
\caption{\yeounoh{A mini-ablation study analyzing the efficacy of  disambiguation, many-shot ICL with synthetic examples for Long Context NL2SQL. The differences (in parentheses) are measured from the full pipeline results in Table~\ref{tab:benchmarks}.}}
\vspace{-0.5mm}
\label{tab:mini_ablation}
\resizebox{\columnwidth}{!}{
\begin{tabular}{l |c c|c c }
\toprule
 & \multicolumn{2}{c|}{w/o disambiguation} & \multicolumn{2}{c}{w/o synthetic examples}\\
 & Ex Acc (\%) & $\bar{T_{g}}$ (T units/req) & Ex Acc (\%) & $\bar{T_g}$ (T units/req) \\
\midrule
SPIDER 1.0 test & 87.1 ( - ) & \cellcolor{red!5}1.3 ( \textbf{$\downarrow$ 0.2}) & \cellcolor{red!5}86.3 ( \textbf{$\downarrow$ 0.8}) & \cellcolor{red!5}1.4 ( \textbf{$\downarrow$ 0.1}) \\
KaggleDBQA test & \cellcolor{red!5}60.5 ( \textbf{$\downarrow$ 0.6}) & \cellcolor{red!5}2.5 ( \textbf{$\downarrow$ 0.2}) & \cellcolor{red!5}60.5 ( \textbf{$\downarrow$ 0.6}) & \cellcolor{red!5}2.3 ( \textbf{$\downarrow$ 0.4})  \\
BEAVER dw & \cellcolor{red!20}52.1 ( \textbf{$\downarrow$ 8.3}) & \cellcolor{red!20}22.5 ( \textbf{$\downarrow$ 467.3})  & \cellcolor{red!0}60.4 ( \textbf{ -}) & \cellcolor{green!10}595.5 ( \textbf{$\uparrow$ 25.5}) \\
\bottomrule
\end{tabular}
}
\end{table}

\begin{table}[t!]
\centering
\caption{\yeounoh{Further breakdown of mini-ablation study results, Ex Acc (\%). As shown in Table~\ref{tab:ablation_study} disambiguation (w/o synthetic examples) is more helpful for challenging or extra hard questions, while many-shot ICL with synthetic examples (w/o disambiguation) is helpful for easier questions.}}
\vspace{-0.5mm}
\label{tab:breakdown_ablation}
\resizebox{\columnwidth}{!}{
\begin{tabular}{l l c c c c }
\toprule
& Question difficulty & Easy & Medium & Hard & Extra Hard  \\
\midrule
\multirow{2}{*}{\begin{tabular}[c]{@{}l@{}} KaggleDBQA  \\ test \end{tabular}} & w/o disambiguation & \cellcolor{green!10}67.56 & 	\cellcolor{green!10}61.62 &	64.86 &	\cellcolor{red!10}43.24 \\
& w/o synthetic examples & \cellcolor{red!10}65.94 & \cellcolor{red!10}57.83 & 64.86 & \cellcolor{green!10}50.81 \\
\midrule
\multirow{2}{*}{\begin{tabular}[c]{@{}l@{}} SPIDER 1.0\\ test \end{tabular}} & w/o disambiguation & 92.09 &	\cellcolor{green!10}89.59 &	87.00 & \cellcolor{red!10}74.19 \\
& w/o synthetic examples & 92.79 &	\cellcolor{red!10}86.79 & 86.79 &	\cellcolor{green!10}75.89\\
\bottomrule
\end{tabular}
}\vspace{-3mm}
\end{table}

Table~\ref{tab:mini_ablation} reports the mini-ablation study, which analyzes the efficacy of the most controversial (expensive) long context techniques - disambiguation and many-shot ICL with synthetic examples - across other benchmark datasets. 
Running the full pipeline without each of the aforementioned techniques reduces both accuracy and latency (or context size). The performance implications of skipping disambiguation (w/o disambiguation) are limited on SPIDER and KaggleDBQA benchmarks, but much more pronounced on the BEAVER dataset. This is because disambiguation is only triggered for executable SQL queries that return empty result sets. Such invalid predictions happens more frequently in the BEAVER data warehouse (dw) dataset, as its schema is significantly more complex and requires more joins (4.25 on the average) compared to other datasets.
Similarly, skipping many-shot ICL with synthetic examples (w/o synthetic examples) reduces both accuracy and latency, but the impact is limited on SPIDER and KaggleDBQA. On the contrary, w/o synthetic examples on BEAVER does not change the accuracy, but also increasing the latency significantly due to more frequent disambiguation attempts. This suggests that, overall, the generated examples are more misleading than helpful, but resulted in fewer empty results. This is expected, as the synthetic example generation primarily targets simpler query structures~\citep{pourreza2024chase} rather than star-schema (data warehouse) queries involving many joins.
It is also interesting to note that disambiguation and synthetic examples have very different performance characteristics. As shown in the full ablation study with BIRD dev (Table~\ref{tab:ablation_study}), disambiguation is more helpful for challenging questions, whereas synthetic examples contribute more to solving simpler problems. We observe the same trends across the other benchmark datasets, as illustrated in Table~\ref{tab:breakdown_ablation}.
\vspace{-3mm}

\section{Discussion and Limitations}

\subsection{Error analysis}
\begin{figure}[t!]
    \centering
        \centering
        \includegraphics[width=0.4\textwidth]{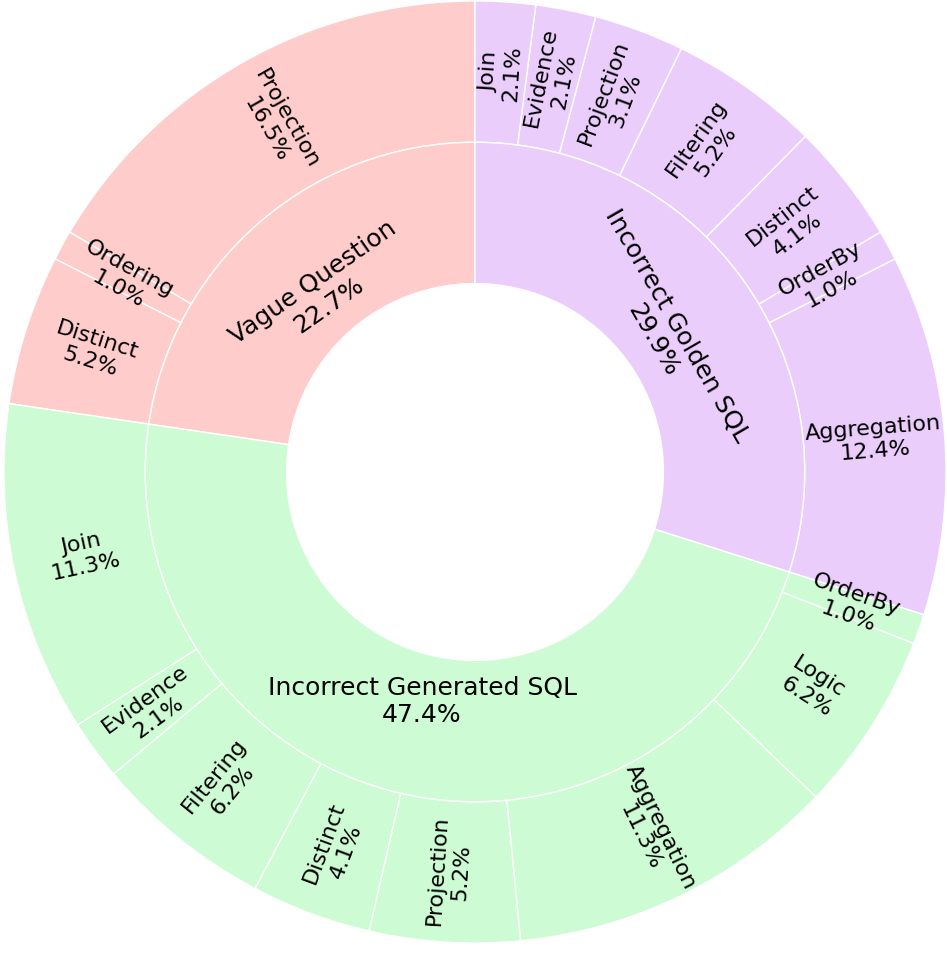} 
    \caption{Breakdown of the observed error categories}
    \label{fig:error_analysis}
\end{figure}

To gain a deeper understanding of the discrepancies between generated SQL and ground truth SQL, we randomly sampled 20\% of the instances where our baseline output deviated from the provided golden SQL in the BIRD dev.  Figure~\ref{fig:error_analysis} presents a breakdown of the observed error categories and their respective proportions within this sampled subset.

The errors are categorized hierarchically. The inner ring of the chart depicts three high-level classifications: "Vague Question" representing instances where the generated SQL was broadly correct but was marked incorrect due to the vagueness of the corresponding NL question;  "Incorrect Generated SQL," indicating cases where the generated SQL contained fundamental errors; and "Incorrect Golden SQL," encompassing scenarios where the provided golden SQL itself is incorrect.

The outer ring provides a finer-grained breakdown of the error types within each high-level category. For example, within "Incorrect Generated SQL," we observe subcategories such as "Join", "Logic", and "Aggregation," each highlighting a specific type of failure in the generated SQL. Specifically, "Logic" refers to cases where the generated query failed to logically understand the intent of the user question. The distribution of errors across these subcategories reveals that issues related to joins, filtering, and aggregation contribute significantly to the overall error rate. 


\subsection{Further performance improvement via self-consistency}
\label{sec:self_consistency}

\begin{table}[t!]
\centering
\caption{Long-context NL2SQL BIRD dev Ex Acc upper bounds (at least one of the candidates is correct) with multiple candidate generation.}
 \vspace{-3mm}
\label{tab:bird_upper_bound}
\resizebox{0.8\columnwidth}{!}{
\begin{tabular}{l c c c c c}
\toprule
\# candidates & 1 & 3 & 5 & 7 & 9 \\
\midrule
Ex Acc (\%) & 65.97 &	68.84	& 68.97	& 69.69	& 70.60\\
\bottomrule
\end{tabular}
}
\vspace{-0.3cm}
\end{table}

Self-consistency is another very popular technique for NL2SQL~\citep{wang2022self,talaei2024chess,pourreza2024chase,dong2023c3,dail_sql,xiyansql}, where multiple candidates are generated, and the most consistent answer or the best one is selected by a fine-tuned picker. \yeounoh{As demonstrated by the recent state-of-the-art approaches~\citep{pourreza2024chase,xiyansql} on the BIRD leaderboard, self-consistency has become a crucial technique for achieving high accuracy in NL2SQL. This involves generating multiple candidates using different generators (a.k.a. NL2SQL pipelines) and each generator trying multiple times, and choosing the best one. 
One of the state-of-the-arts, \textit{CHASE-SQL}, in Table~\ref{table:bird_dev_results} uses our long context pipeline as one of the three candidate generators, contributing to its top-tier performance on the BIRD leaderboard.
One caveat of self-consistency is that it can quickly become expensive in terms of latency and the number of LLM calls. Self-consistency is orthogonal to our work, and can further improve the NL2SQL performance if used together with our pipeline. Table~\ref{tab:bird_upper_bound} illustrates the impact of generating multiple output candidates from our long context pipeline using a generation temperature of 0.5. With an oracle picker (i.e., at least one of the candidates is correct per question) we can uplift the accuracy beyond 70\%. In \citep{pourreza2024chase,xiyansql}, the candidate sets are generated using multiple pipelines, as different strategies result in more diverse outputs, further boosting the final accuracy of self-consistency.}

\yeounoh{
Self-consistency is orthogonal to the use of long context and falls outside the scope of this paper.
In this work, we focus on generating one high quality candidate and study the performance implications of leveraging long context (i.e., extra contextual information), that has not been previously explored for NL2SQL.}
\vspace{-2mm}
\subsection{Long context in production and limitations}\label{sec:production_limitation}
Our study reveals (Table~\ref{tab:ablation_study}) that a good chunk of requests (roughly 68\%) from BIRD dev can reach ~400k accumulated tokens and take up to ~130 $T$ units. This would likely be prohibitive for most production services. And the cost will be even higher with techniques like self-consistency. Long context is a great means to add more appropriate information, but it has to be used with some care. To leverage long context properly in production, one may pick and choose which extra information and techniques are appropriate based on our analysis. For instance, one can skip disambiguation and online example generation to bound the latency of the majority of requests to at around <1.5 normalized seconds. The example can be generated offline, and as shown in Section~\ref{sec:synthetic_example_selection}, example selection works well with synthetic examples. Using \textit{flash} can further reduce the cost, roughly -20\% in latency and -94\% in dollars per request given the current pricing and the observed latency.

\yeounoh{One limitation of our study, as we analyze the impact of long context and additional contextual information on NL2SQL, is the absence of benchmark datasets that accurately model enterprise use cases. Such cases typically involve hundreds of very wide tables with thousands of columns, as well as schemas with less descriptive and similar table and column names, making schema linking significantly more challenging. These scenarios not only pose inherent difficulties for schema linking, but further exacerbate the cost of long-context techniques due to the increased amount of schema information. 
To address this, we incorporated multiple challenging NL2SQL benchmark datasets and evaluated how the long-context model performs in cases with very low density information (e.g., providing full schema details and sample values from tens of irrelevant tables and DBs, while only a few columns are relevant, as shown in Table~\ref{tab:tbr_bird_ex_acc}). 
We observe that extra information is generally not distracting, and the long context techniques perform as well as the baselines without them. While filtering and ranking the most relevant information remain crucial for both accuracy and cost, our findings suggest that, unless retrieval is highly accurate, increasing recall by including more information - even at the expensive of lower precision - is a beneficial strategy. We expect that these insights will hold for more extreme use cases and become increasingly evident as long context LLMs continue to improve. For instance, our long-context NL2SQL pipeline yields significantly better results on the enterprise data warehouse benchmark BEAVER compared to the baseline results from~\citep{chen2024beaver} (see Table~\ref{tab:benchmarks}).
}

While this work focuses on evaluating the impact of providing various forms of contextual information directly within the LLM's extended context window, we acknowledge that the explored information is not exhaustive. Notably, we did not investigate the utilization of knowledge graphs for enhancing NL2SQL performance. While it is expensive to build a comprehensive knowledge graph for a given domain, recent works~\citep{allemang2024increasing,sequeda2024benchmark} have shown promising results using knowledge graph-based question answering over enterprise datasets. In future work, we are interested in exploring the integration of additional contextual sources, such as business ontologies and knowledge graph, to further leverage the capabilities of long-context LLMs.
\vspace{-3mm}

\section{Conclusion}
\yeounoh{
In this work, we explored the potential of leveraging the extended context window offered by Google's \textit{gemini-1.5-pro} for NL2SQL. Our findings demonstrate that long-context LLMs can effectively utilize the additional context, achieving strong performance across multiple benchmarks (Table~\ref{tab:benchmarks}), including 67.41\% execution accuracy on the BIRD benchmark - without fine-tuning or computationally expensive self-consistency techniques. 
This performance highlights the robustness of the long-context models in retrieving and reasoning over extensive contextual information. Specifically, we analyze the impact of including entire DB schema details, user provided hints, sample column values, and synthetic examples for many-shot ICL to improve SQL generation accuracy. We also show that self-correction and verification can further enhance accuracy, although at the cost of increased latency and accumulated tokens.
Overall, we observe that \textit{gemini-1.5-pro} exhibits strong retrieval capabilities over the extended context window, even in the presence of irrelevant information. Additionally, our findings indicate that the long context models are more robust and do not suffer from the ``lost in the middle''~\citep{liu-etal-2024-lost}) issue.
}

\yeounoh{
Our study is a unique example of how enhanced capabilities of LLMs, in this case the extended large context size, impact the way we approach the NL2SQL problem. In contrary to prior works that focused on squeezing the filtered information into a limited context window, we explored the potential of providing additional useful information to the model. We showed that the additional information can be useful in NL2SQL by helping to resolve semantic issues (SQL syntactic issues for the common SQL dialect are already well addressed by the base model). Furthermore, we investigated the cost implications of using the long context techniques and concluded that they can be complementary and more efficient when combined with accurate schema and example retrievals, with an emphasis on recall. 
In fact, perfect table schema retrieval would yield stronger performance  (Appendix~\ref{appendix:impact_col_selection}) by narrowing the schema linking search space during SQL generation~\citep{chen2024beaver,Floratou2024NL2SQLIA,talaei2024chess}. However, achieving perfect retrieval is highly challenging in practice (Appendix~\ref{appendix:tbr_simulation}).
}

\yeounoh{
In real-world scenarios where the retrieval and ranking are sub-optimal, providing more information and relying on the long context models' strong retrieval capability can offer a viable, though more expensive, alternative strategy. Improving the cost efficiency of long context model serving would be an important area of research to make long context more practical.
}
\vspace{-2mm}
\begin{acks}
We would like to thank David Culler, Hank Levy, Reza Sherkat, Per Jacobsson, Cosmin Arad, Zeke Miller, Eva Sharma,  Richard Kuzma,  Xiance Si, Zhixian Yan and the anonymous reviewers for their valuable feedback.
\end{acks}

\balance
\bibliographystyle{ACM-Reference-Format}
\bibliography{main}

\appendix
\section{appendix}

\subsection{Table retrieval simulation}
\label{appendix:tbr_simulation}
For a realistic table retrieval simulation, we ran the BIRD bench against BigQuery's SQL Code Assist feature\citep{google_code_assist}. The result differs from the standard BIRD bench setup, as it mirrors the challenges of serving production users of BigQuery. Conventionally when running the BIRD bench, each query is linked to a single database containing up to 13 tables. BigQuery operates differently; users may ask questions about any table that they have access to, and BigQuery's retrieval system uses user interaction histories to reduce the search space. For realistic results the user interactions were seeded to match the distribution of queries in BigQuery's production traffic, with a bias towards session-starting queries (less repeated queries), as these present a more difficult retrieval problem.

For each example in the benchmark, table retrieval operates as follows: 1) The simulated user view and query interactions narrow the search space down to ~1-100 candidates; 2) The candidates are embedded and re-ranked with Gecko \citep{Lee2024GeckoVT} for relevance to the user query; 3) A fixed top-k candidates are passed on to the NL2SQL stage.
At the time of test, the recall of the first stage was 82\%. As such, 82\% is the maximum attainable end-to-end recall, and represents a near perfect result from the re-ranking stage.




\subsection{Impact of good column selection}
\label{appendix:impact_col_selection}

\begin{table}[ht!]
\centering
\caption{Relevant schema element filtering (TBR: Table Retrieval, CR: Column Retrieval) and Ex Acc on BIRD dev}
\label{tab:columnSelection}
\vspace{-3mm}
\resizebox{0.9\columnwidth}{!}{
\begin{tabular}{l c c}
\toprule
& Filtered Schema & Ground Truth Schema \\
\midrule
Ex Acc (\%) & 64.08 & 72.43 \\
TBR Recall (\%) / Precision(\%)  & 97.69 / 89.72 & 100 / 100 \\
CR Recall(\%) /  Precision(\%) & 97.12 / 69.43 & 100 / 100 \\
\bottomrule
\end{tabular}
}\vspace{-2mm}
\end{table}

We evaluate the impact of schema selection on execution accuracy using the BIRD dev. We compare a filtered schema against a perfect ground truth schema.  The filtered schema, constructed similarly to CHESS~\citep{talaei2024chess}, incorporates relevant tables and columns, their descriptions, and relevant example values.
~\ref{tab:columnSelection} reports execution accuracy (Ex Acc), as well as table retrieval (TBR) and column retrieval (CR) recall and precision. While the ground truth schema achieves perfect performance (100\%) on all metrics, the filtered schema demonstrates high recall, correctly identifying most relevant schema elements (97.69\% table retrieval recall and 97.12\% column retrieval recall). However, precision is lower (89.72\% for table retrieval and 69.43\% for column retrieval), indicating the inclusion of some irrelevant columns/tables.  
It is important to note that the schema filtering algorithm incurs substantial cost, requiring an average of 78 LLM calls and 339,965 tokens per request on the BIRD dev.

\end{document}